\documentclass[aps,prb,floatfix,twocolumn,showpacs]{revtex4-1}
\usepackage{graphicx}
\usepackage{dcolumn}
\usepackage{bm}

\begin{document}

\title{Strain-gradient-induced switching of nanoscale domains in free-standing ultrathin films}
\author{G. D. Belletti, S. D. Dalosto, Silvia Tinte}
\affiliation{Instituto de F\'{i}sica del Litoral-CONICET,
Universidad Nacional del Litoral,
G\"{u}emes 3450, (3000) Santa Fe, Argentina}

\date{\today}

\begin{abstract}
We report first-principle atomistic simulations on the effect of local strain gradients 
on the nanoscale domain morphology of free-standing PbTiO$_3$ ultrathin films. 
First, the ferroelectric properties of free films at the atomic level are reviewed.
For the explored thicknesses (10 to 23 unit cells),
we find flux-closure domain structures whose morphology is thickness dependent.
A critical value of 20 unit cells is observed: 
thinner films show structures with 90$^\circ$ domain loops, 
whereas thicker ones develop, in addition, 180$^\circ$ domain walls, 
giving rise to structures of the Landau-Lifshitz type.  
When a local and compressive strain gradient at the top surface is imposed,
the gradient is able to switch the polarization of the downward domains, 
but not to the opposite ones.
The evolution of the domain pattern as a function of the strain gradient 
strength consequently depends on the film thickness. 
Our simulations indicate that in thinner films, 
first the $90^o$ domain loops migrate towards the strain-gradient region, 
and then the polarization in that zone is gradually switched.
In thicker films, instead, the switching in the strain-gradient region 
is progressive, not involving domain-wall motion, 
which is attributed to less mobile $180^o$ domain walls.
The ferroelectric switching is understood based on the knowledge of the 
local atomic properties, and the results confirm that mechanical flexoelectricity provides 
a means to control the nanodomain pattern in ferroelectric systems.
\end{abstract}


\maketitle

\section{Introduction}
Flexoelectricity, the linear coupling between electric polarization $P$
and inhomogeneous deformation, is always allowed by symmetry
in all insulators since strain gradients break inversion symmetry.
This effect, found 50 years ago~\cite{kog64}, has traditionally received 
little attention owing to its relatively weak effect in macroscopic samples.
At the nanoscale, however, strain gradients are typically much larger  
than at macroscopic scales, as reduced dimensions imply larger gradients.
The impressive progress in the design and control 
of nanoscale structures has motivated a revival interest in flexoelectricity,
which has become an attractive topic in material science
in the past few years~\cite{zub13,yud13,lee12}.
In the context of perovskite thin films, recent experimental 
breakthroughs~\cite{leePRL11,gruv03,luSCI12} have revealed that the effect of 
strain gradients can be enormous in thin films, 
large enough to rotate or even switch ferroelectric domains. 
For instance, a recent experimental work~\cite{luSCI12}, which in part motivated 
ours, shows that an inhomogeneous deformation caused by pressing the tip of
an atomic force microscope over the surface of BaTiO$_3$ films is able to 
switch the electric $P$.

To model in a more realistic manner the flexoelectric switching in ultrathin 
ferroelectrics, it is important to include the domain morphology, which is expected 
to depend strongly on the mechanical and electrical boundary conditions~\cite{yud13}.
In the absence of electrodes or other sources of free charges that compensate   
the depolarizing field, it is well known that ferroelectrics respond by 
forming periodic arrays of $180^o$ domains, whose character might be 
described using the traditional models of Kittel (open $P$ flow) 
or Landau-Lifshitz (closed $P$ flow).
For instance, ordered $180^o$ stripe domains of the Kittel type
have been observed in PbTiO$_3$ (PTO) films as thin as 3 unit cells 
deposited on SrTiO$_3$ substrates, using x-ray photoelectron diffraction~\cite{fong04}. 

From the theoretical side, 
to understand and predict properties of nanoscale films at the atomistic level,
there have been numerous studies using $ab initio$ methods 
as well as simpler models constructed from them, 
such as model Hamiltonians and atomistic core-shell models. 
In particular, in PTO ultrathin films the ferroelectric and antiferrodistortive 
distortions in the (001) surface have been described~\cite{mey02,bung05,sepli0506}, 
indicating also a decrease of tetragonality with stabilization 
of in-plane $P$ and antiferrodistortive reconstruction in the region near the surface. 
Studies of domain morphology reported vortex-like structures in films
of a few unit cells thick~\cite{lai07,behe08}.
The changes in $P$ and lattice deformation across the $90^o$ and $180^o$ domain walls (DWs) 
hace also been investigated, with calculations performed in the material bulk~\cite{mey02,shim08}
and in films of 3-unit-cell thickness~\cite{shim10,qu13}.
However, none of these works describe the local properties of PTO ultrathin films and
their correlation with the domain morphology as the thickness varies.
Concerning this kind of simulations treating flexoelectricity, we can cite
recent simulations involving a modified model Hamiltonian technique 
for (Ba$_x$Sr$_{1-x}$)TiO$_3$~\cite{ponoPRB12} which has been used to calculate 
flexoelectric coefficients as a function of the film thickness and temperature, 
but without making any reference to the domain presence.

Then, although the flexoelectricity effect on nanoscale is attracting much attention, 
its exact role in the poling process of ultrathin films with a nanodomain 
configuration has not been completely characterized from the atomistic level.
In this work, we address some particular questions, such as How does a local strain 
gradient affect the domain pattern in free-standing 
ultrathin films? and How does this depend on the film thickness and the strain 
gradient strength?
To answer these, using a first-principles-based atomistic model 
we study the flexoelectric switching in ultrathin films of 
the stereotypical ferroelectric PTO, when a longitudinal strain gradient 
is applied locally over the center of one domain.
Our simulations indicate that for the explored films, thinner than 10~nm,
the domain patterns in the unstrained films are thickness dependent
and responsible for the film being locally deformed.
In agreement with this, the film response to a local and 
longitudinal strain gradient varies according to the 
nature of the domain pattern and as a function of the gradient intensity. 
The paper is organized as follows.
Section II describes the computational details for modelling the film
and the conditions for mimicking a strain gradient.
Section III presents the results in two parts.
First, 
the local properties of strain-free films are reported 
for two thicknesses in order to understand 
later the gradient-induced changes.
Second, the domain pattern evolution for each of these films
is investigated as the strain gradient strength is enhanced.
Finally, the paper concludes with a summary in Sec.IV.

\section{Methodology: computational details}

Prototype ferroelectric PTO is modeled using an atomistic core-shell model 
fitted completely to first-principles calculations 
(i.e., no explicit experimental data were used as input)
as described in detail in ref.~\onlinecite{sepli04}.
The model contains fourth-order core-shell couplings, 
long-range Coulombic interactions, and short-range
interactions.
It has shown to correctly reproduce  
the cubic-tetragonal phase transition in bulk~\cite{sepli04}
and also was able to describe the surface properties and
interface effects of ultrathin films~\cite{sepli0506}.

Here we study free-standing ultrathin films of PTO under 
ideal open-circuit electric boundary conditions
performing classical molecular dynamics (MD) simulations using
the \textsc{DL$_-$POLY} package~\cite{dlpoly}
at a constant temperature of 50~K,
where PTO presents a tetragonal ferroelectric phase.
Our films have two PbO-terminated surfaces and are modeled 
using supercells of $N_x\times N_y\times N_z$ unit cells.
Periodic boundary conditions are applied along the in-plane 
pseudocubic [100] and [010] directions (axes $x$ and $y$, respectively)
but not along the [001] one (axis $z$) defined as the growth direction.
The long-ranged electrostatic energy and forces are calculated 
by a direct sum method~\cite{wolf}
which is a computationally efficient alternative to the Ewald summation.
In the $xy$ plane, the simulated supercell is kept constant
and constrained to form a simple square lattice ($N_x = N_y$)
of edge $N_x\times a$ with $a$=0.3866~nm,
the in-plane lattice parameter in bulk 
computed at 50~K with the present model.
The film thicknesses range from 4.0~nm (10$c$)
to 9.2~nm (23$c$), where $c$ is the lattice parameter in the growth direction.
Atoms are free to relax along all directions, 
except when mentioned, as long as they remain inside the simulation box.

Local polarization is defined here as the dipole moment 
per unit volume of a perovskite unit cell centered on the Ti ion
and delimited by the eight Pb nearest neighbors at the corners of the box. 
We consider contributions from all atoms in the conventional cell and
atomic positions with respect to that center:
\begin{equation}
{\bf p} = \frac{1}{\it v} \sum_{\it i} {\frac{1}{\omega_{\it i}} {\it q_i} (\bf r_{\it i} - \bf r_{\it Ti} )}
\label{eq:pol}
\end{equation}
where $v$ is the volume of the cell, 
$q_i$ and $\bf r_{\it i}$ denote the charge and the position of the $i$ particle, respectively, 
and $\omega_i$ is a weight factor equal to the number of cells to which the particle belongs. 
The reference position $\bf r_{\it Ti}$ corresponds to the core of the Ti atom, 
and the sum runs over 29 particles, including cores and shells of the 
surrounding ions (8 Pb and 6 O atoms) plus the shell of the Ti atom. 
Note that $\bf p$ is independent of the origin for the $\bf r_{\it i}$ vectors 
and vanishes when atoms are at their high-symmetry positions.
The charge values $q_i$ are given in Ref.~\onlinecite{sepli04}.

\begin{figure}
\begin{center}
\includegraphics[width=6.cm,angle=0]{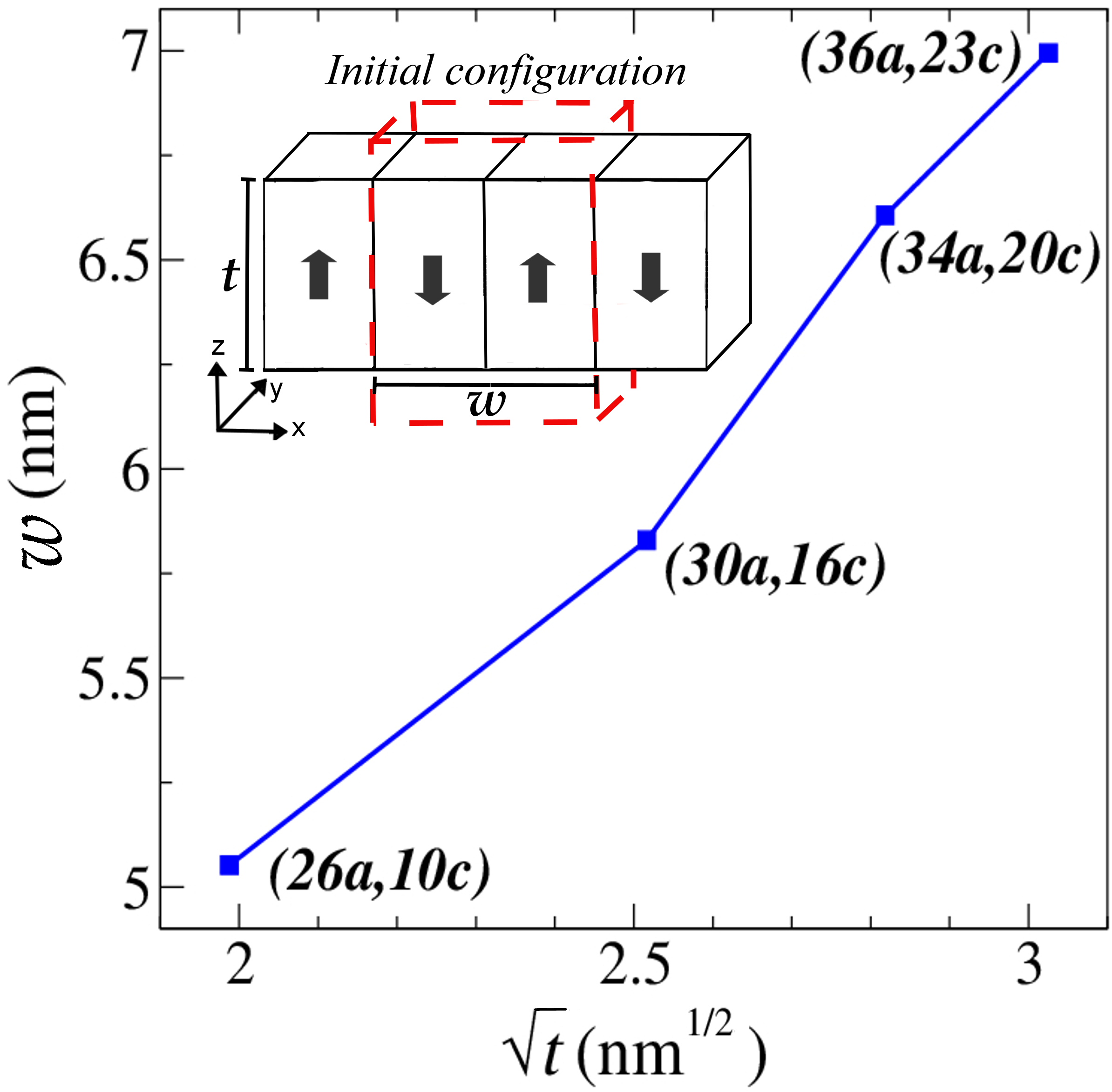}
\end{center}
\vspace{-15pt}
\caption{Optimal width $w$ of the simulation supercell 
for films of different thickness $t$. 
For each film, $w$ and $t$ are informed by the pairs 
($N_x a$, $N_z c$) that label the supercell size along $x$ and $z$,
respectively. \\
Inset: Schematic of the initial configuration of domains
used to start the MD simulations for all the thicknesses.
Supercell boundaries are indicated by dashed lines.}
\label{fig:kittel}
\end{figure}

The initial configuration of the strain-free film
for all the considered thicknesses 
is set up by creating 180$^\circ$ stripe domains 
with polarizations pointing upward and downward normal to the surface, 
alternated along the $x$ axis, 
as schematically shown in the inset of Fig.~\ref{fig:kittel}. 
The simulation supercell, indicated by dashed lines, holds only two domains. 
Based on previous results~\cite{poy99,mey02} and also ours~\cite{belle}, which 
indicate that the 180$^\circ$ DWs centered at the Pb-O lattice planes
are more stable than those at the Ti-O$_2$ ones,
we use the former in our starting configuration.
Note that letting the structure relax will allow this domain configuration 
to change and the optimized one will depend on the film thickness
as described below.

\begin{figure*}
\begin{center}
\includegraphics[width=17.0cm,angle=0]{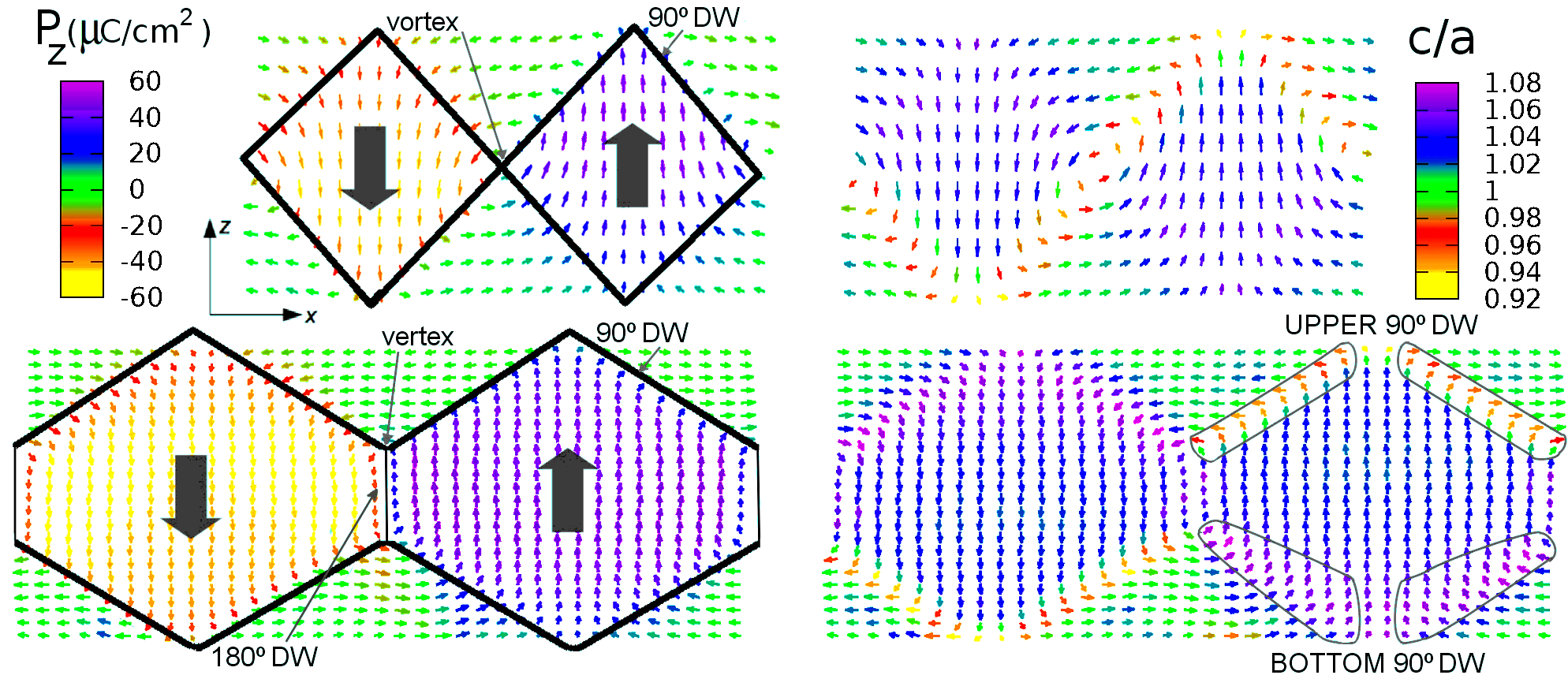}
\end{center}
\vspace{-15pt}
\caption{(Color online) Cell-by-cell polarization map projected on the $xz$ plane 
for strain-free films of 10$c$ (top panels) and 
23$c$ (bottom panels) thickness.
Colors represent (left) the polarization $z$ component ($\mu$C/cm$^2$) and 
(right) the $c/a$ tetragonal relation. 
Domains are delimited by solid lines: thick and thin for 90$^\circ$ and 180$^\circ$ domain walls, respectively.
Large arrows label the UP ($+P_z$) and DOWN ($-P_z$) domains. In the bottom-right panel, the two types
of 90$^\circ$ domain walls are indicated.}
\label{fig:pol_free}
\end{figure*}

{\it Strain gradient description.} As we are interested in the effect 
of switching the out-of-plane polarization, 
we apply strain gradients in the direction normal to the surface, 
that is, a gradient of the out-of-plane strain component $\eta_{zz}=c/c_0 - 1$ 
along the [001] direction denoted as $\partial\eta_{zz}/\partial z$.
We impose a linear gradient localized on an in-plane square region 
of $4\times4$ unit cells that penetrates until the central plane,
which compresses the upper plane by a given strain $-\eta_{zz}^{max}$
and then decreases linearly in the intermediate planes 
until vanishing at the central cells.
One advantage of using the present atomistic model to treat flexoelectric 
effects~\cite{flex} is that the interatomic potential does not need to be modified; 
it is sufficient to place the system under appropriate
mechanical-boundary conditions that mimic the required gradient.
To do that, one first has to establish the strain gradient in the film,
noting that its presence implies a stress gradient which produces 
net forces acting on the atoms inside each unit cell.
During MD simulations,
the atoms will try to relax in order to vanish these net forces.
Thus, in order to preserve the gradient
a force pattern must be applied to the atoms in the unit cell;
i.e., certain atoms must be fixed along the full simulation.
Specifically, to set up the gradient, we start from the relaxed free film 
(which already presents a domain structure)
and displace rigidly the atomic planes of the chosen unit cells until 
the average distance between consecutive Pb-O lattice planes 
equals the required local strains $\eta_{zz}$
in order to have $\partial\eta_{zz}/\partial z$.
We choose to freeze out the $z$ components of 
the Pb-core positions, allowing all other internal degrees of freedom (the
Pb shell and other atoms, cores plus shells) to relax along all directions.
Other choices are possible, for instance, 
the force pattern could be distributed between 
all atoms in the unit cell in different proportions~\cite{DHV_11}.
Finally, note that the local strain gradient induces an internal electric field 
which can compensate the depolarization field
allowing that part of the free-standing film is poled. 

\section{Results}

\subsection{Free-standing ultrathin films of PbTiO$_3$}

We focus first on strain-free films under ideal open circuit boundary 
conditions and revisit their local features.
To begin with, the optimum lateral size of the simulation supercell 
(i.e., the number of cells $N_x$) is determined.
For each considered thickness, we build up supercells of different lateral sizes, 
relax their structures, compare their total energies, and choose the lowest-energy one.
The optimized structures are described below; here we just 
present the optimum width $\omega$ of one domain period as a function 
of the film thickness $t$.
Figure~\ref{fig:kittel} displays the linear relation between $\omega$ 
and the root square of $t$,
showing the fulfillment of the Kittel law.
We found that the optimum lateral sizes of the supercells are 26$a$, 30$a$, 34$a$,
and 36$a$ for thickness of 10$c$, 16$c$, 20$c$, and 23$c$, respectively.

Hereafter we present results only for films of two thicknesses — 
4.0~nm and 9.2~nm — because of their representative domain structures.
Figure ~\ref{fig:pol_free} shows a side view of cell-by-cell 
polarization map for the two thicknesses, 
where colors indicate the $z$-component values in the left panels 
and the $c/a$ cell tetragonality in the right ones.
Starting with the left panels,
the 10$c$ ultrathin film (top panel) shows a flux-closure arrangement 
around central cores or {\it vortices}.
This vortex state, 
where the local $P$ orientation rotates continuously around one geometrical core,
can also be described as a 90$^\circ$-domain loop.
Indeed, we can demarcate upward and downward $P$ domains 
(hereafter UP and DOWN, respectively) 
with a rhombus shape surrounded by triangular caps 
(where $P$ is almost parallel to the surface)
that complete an overall closed loop. 
Our simulations show that this domain morphology holds for thicker films
until thicknesses of $\sim$~20 unit cells.
Above that critical value, 
the central vortices enlarge in the normal 
direction, giving rise to 180$^\circ$ DWs 
that grow together with the film thickness.
Consequently, the shape of the UP and DOWN domains changes 
to hexagonal and a domain structure as envisioned by Landau-Lifshitz is formed,
as shown in the bottom panels for a 23$c$-thickness film.
In this structure, in particular, we see that the 180$^\circ$ DWs end up 
in {\it vertices} or confluences of three DWs
as predicted by Srolovitz and Scott~\cite{scott86} 
and recently confirmed with high-resolution transmission
electron microscopy images of Pb(Zr$_{1-x}$Ti$_x$)O$_3$~\cite{jia11}.

As regards the thicknesses of the DWs, 
they vary depending on the type and location inside the film, 
the difference being more evident in the thicker films.
In the inner region, the orientation of the $P$ changes abruptly
from up to down over a distance of two lattice constants,
creating very narrow 180$^\circ$ DWs, in agreement
with previous theoretical results for defect-free walls~\cite{mey02}.
The 90$^\circ$ DWs instead differ according to their position with 
respect to the out-of-plane $P$ orientation, as indicated schematically 
in the bottom-right panel in Fig.~\ref{fig:pol_free} for the UP domain
as an example, but the same behavior is repeated in the DOWN domain.
In the top part, the $P$ rotates 90$^\circ$ in only three unit cells, 
leading to narrow 90$^\circ$ DWs. 
In the bottom part, instead, the $P$ rotates more continuously, requiring 
a few unit cells to change from the in-plane $P$ at the surface until 
pointing perpendicular, making these 90$^\circ$ DWs much broader.
This behavior can be attributed to the electrostatic repulsion between 
the head-to-head~\cite{wu06,gur11} in-plane $P$s that confront each other 
at the junction of two triangular domains. 

The cell-by-cell tetragonal $c/a$ ratio is indicated by colors in
the right panels of Fig.~\ref{fig:pol_free},
where $c$ and $a$ were obtained by taking the distances between 
pairs of nearest-neighbor Pb atoms along $z$ and $x$, respectively.
The cells inside the triangular domains with $P$ parallel 
to the surfaces are almost cubic,
whereas those at the cores of the domains with $P$ perpendicular
are tetragonal and their values are thickness dependent. 
Further, and in agreement with the high strain-$P$ interaction in PTO,
these tetragonality maps demonstrate the difference between the 90$^\circ$ DWs.
The narrow upper ones are well defined,
presenting an abrupt change in the $c/a$ ratio
as the tetragonal axis rotates from perpendicular to
parallel to the surface;
however, the broader ones at the bottom are characterized by a large tetragonality 
(reaching, for instance, $\sim$8\% in the 23$c$ film)
which attenuates gradually in a few unit cells towards both sides.
Concerning the 180$^\circ$ DWs in the thicker films, 
the tetragonality does not change notably,
rather the wall is defined by an abrupt relative displacement or {\it offset}
of the Pb atoms along $z$.
Finally, we emphasize the significant lattice deformation 
observed at the surfaces (shown by different colors in the right panels of Fig.~\ref{fig:pol_free})
as well as across the film thickness specially in the 90$^\circ$ DWs.

\begin{figure}
\begin{center}
\includegraphics[width=7.5cm,angle=0]{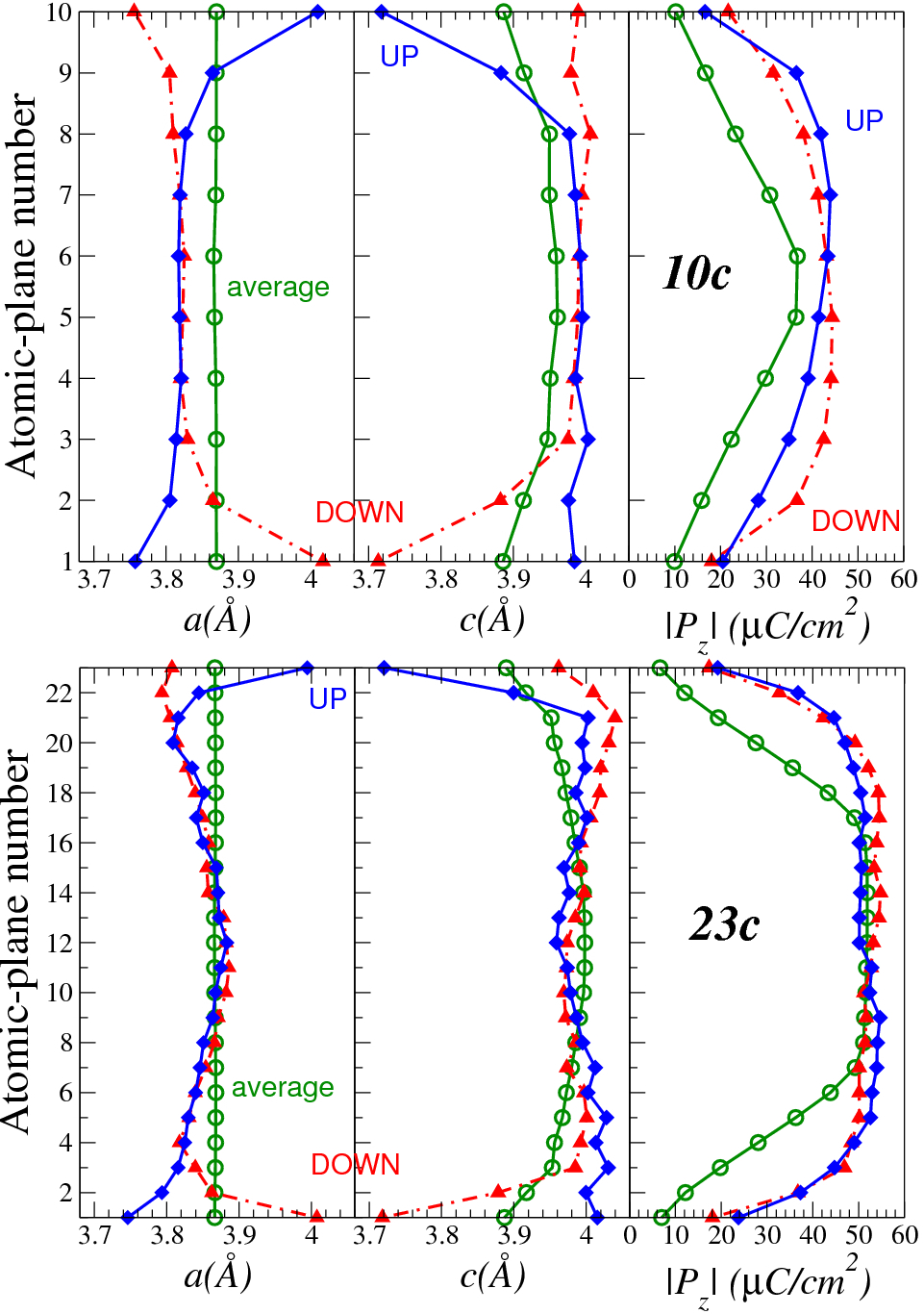}
\end{center}
\vspace{-15pt}
\caption{(Color online) Local lattice parameters $a$ and $c$ and
modulus of the polarization $z$-component across the thickness for the 10$c$ and 23$c$ films.
Local parameters are averaged at the central cells of each domain —UP [filled (blue) 
diamonds] and DOWN [filled (red) triangles]—and over the complete $xy$ planes 
[open (green) circles].}
\label{fig:latt_free}
\end{figure}

To obtain further insights, 
we monitor those properties in detail through the film thickness.
Figure ~\ref{fig:latt_free} shows the in-plane $a$ and out-of-plane $c$ 
lattice parameters across the thickness for both films.
Averaging these values over complete planes parallel to the surface [open (green) circles], 
we find that the average local $a$'s remain constant across the film
due to the mechanical boundary conditions imposed on the simulation supercell.  
The average $c$ values show the typical symmetric profile
determined by the two free surfaces: they are compressed at both surfaces 
and gradually increase towards the inner film,
consistent with previous results~\cite{sepli0506}.
However, examining closely at the central region of each domain,
the local $c$'s present an asymmetric distribution:
they are highly compressed in the two unit cells of the top surface 
(with respect to the $P$ orientation)
and slightly enlarged at the bottom ones. 
The strain change along [001] is compensated in the plane
only by the local lattice parameters $a$, which behave oppositely to $c$,
but not by the $b$ parameters, which remain almost constants (not shown here),
corroborating that the unit cells at the surface neighborhood are 
highly distorted.
This particular asymmetry through the thickness was also predicted 
in superlattices of PTO/SrTiO$_3$ from first-principles calculations~\cite{junq12}
and even in the limit case of PTO films of 1$c$ thickness  
whose unit cell is distorted~\cite{qu13}. 
Thus, this information together with the above $c/a$ description
shows that the free surfaces induce spontaneous strain gradients 
that break the inversion symmetry in the film,
in agreement with previous works~\cite{shim10,qu13}.
Furthermore, knowledge of the local strain maps 
is essential to predict and understand the effect of 
an imposed strain gradient. 
\\
Moving to the right panels of Fig.~\ref{fig:latt_free},
the profiles of the local $P_z$ modulus
show that the averaged values per lattice plane are symmetric.
Remarkably, inside each domain, 
opposite to the highly asymmetric profiles of local $c$,
the $P_z$ profiles are almost symmetric
(in thinner films, actually they are slightly enhanced towards the top part).
Then we note that the modulus of $P$ and not $P_z$ follows the $c$ behavior across the film.

\begin{table}[ht]
\caption{Oxygen octahedra rotations (in degree) of the $AFD_z$
distortions for two thicknesses ($t$) in different atomic layers
starting at the top surface, given at the center of the UP, DOWN and triangular domains,
and averaging per full TiO$_2$ atomic layer. 
$Ab initio$ results after Ref.~\onlinecite{bung05} are
reported in parenthesis.
The last row of data corresponds to our calculations performed in the bulk
material with two domains separated by 180$^\circ$ domain walls. } 
\centering 
\begin{tabular}{c c c c c c} 
\hline\hline 
$t$ & Atomic layer &  UP  & DOWN & Triangular & Average \\[0.5ex] 
\hline 
10$c$ & 1 &   14.2 & 13.3  & 13.0 &  13.0 (11.4)  \\ 
      & 2 &   1.0  &  2.9  & 3.5  &  2.8 (2.9)   \\
      & 3 &   6.0  &  6.2  & 6.4  &  6.5 (3.9)  \\
      & 4 &   4.9  &  5.1  & 5.2  &  4.9    \\
      & 5 &   4.8  &  5.0  & 5.8  &  5.2   \\
\hline
23$c$ & 1  & 14.3 & 13.0 & 13.0 & 13.1 (11.4)\\ 
      & 2  & 0.9  & 2.8  & 4.1  & 3.0 (2.9)  \\
      & 3  & 5.7  & 6.3  & 6.5  & 6.6 (3.9)  \\
      & 4  & 3.1  & 4.3  & 6.4  & 5.16    \\
      & 5  & 3.0  & 4.1  & 6.1  & 5.1  \\
      & 11 & 0.8  & 1.0  & 4.0  & 1.6   \\
\hline 
$bulk$  &  & 1.0  & 1.0  & 3.9  & 1.0 (3.3) \\
\end{tabular}
\label{table:AFD} 
\end{table}

To end up this section, we analyze the local distribution 
of the antiferrodistortive distortion
which has been shown~\cite{bung05, step01}
to coexist with the polar distortion at the surface, 
eventhough it essentially vanishes in bulk.
It involves the oxygen octahedron rotation around the central Ti ion, where
the rotation axis is the $z$ axis ($AFD_z$)
with out-of-phase movements between neighbor cells on the (001) plane but
successive in-phase movements along the [001].
Its strength is described by the rotation angle of TiO$_4$ 
in the normal layer at each unit cell.
Table~\ref{table:AFD} lists the $AFD_z$ rotation angle dependence 
of the atomic layers starting at the top surface,
for the two films of interest and in different regions of the film 
by entering perpendicular to the surface:
in the central cells of the  UP and DOWN domain,
and crossing the center of one triangular domain
(reaching the central vortex of the 90$^\circ$ DW loop 
or the 180$^\circ$ DWs, depending on the film thickness).
The last column reports the values averaged for each complete TiO$_2$ atomic layer
parallel to the surface.
Moving along [001] in the center of one domain,
it is found that the rotation angle of $\sim 13^o$ 
in the superficial cells decays towards the internal atomic layers: 
although it almost vanishes in the second layer,
recovers in the third one, and then keeps decreasing slowly.
It reaches the bulk value of $\sim$ 1$^o$ just 
at the eleventh layer of the 23$c$~film,
while in the thinner film, the superficial effects predominate.
Furthermore, we note that cells in opposite surfaces differ in $\sim 1^o$, 
which is also related to the different natures of the junction of the 90$^\circ$ DWs
as explained before.
On the other hand, when entering the film through the triangular domains,
the $AFD_z$ distortions from the second layer towards the inner ones 
are a bit larger than those in the center of the UP and DOWN domains.
Specifically, in the 180$^\circ$ DWs the $AFD_z$ distortion
is about $4^o$ (the octahedra rotate to allow the abrupt $P$ change,
in addition to the offset), versus the $\sim 1^o$ in the domain cores, where the cells 
are tetragonal with ferroelectric distortions mostly along $z$.
The averaged values show a behavior similar to that in the cores
of the out-of-plane $P$ domain but with values $\sim$ 1$^o$ larger
than in the inner cells due to the values in the DWs.
Finally, note that our results for the first three atomic layers  
can be compared, and indeed fully agree with previous calculations reported using  
the present model~\cite{sepli0506} as well as first-principles methods~\cite{bung05}.
Although those calculations have provided evidence of $AFD_z$ variation 
in the near-surface region, 
they were performed using smaller supercells than here and
without any reference to the effect of the domain pattern.
Hence, our inspection of the $AFD_z$ local distribution 
gives insights into the dependence of the domain morphology,
especially in the DW neighborhood. 

\subsection{Film response to a local strain gradient}

\begin{figure*}
\begin{center}
\includegraphics[width=18cm,angle=0]{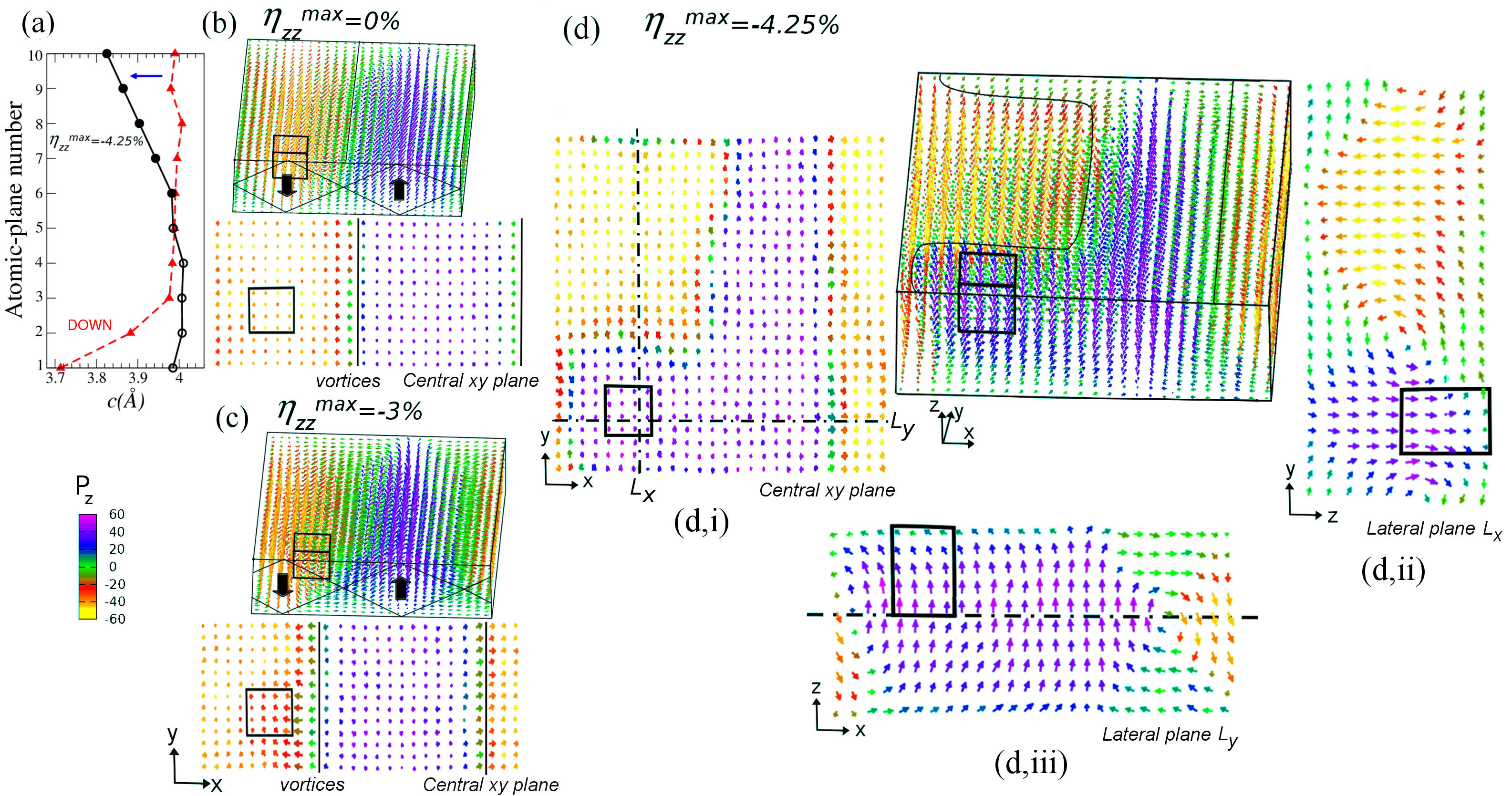}
\end{center}
\vspace{-15pt}
\caption{(Color online) 
Results for a 10$c$-thickness film. 
(a) Local lattice parameter $c$ in the central region of the DOWN domain  
for a strain-free film (triangles) 
and after the strain gradient is applied (circles).
(b-d) Cell-by-cell $P$ distributions for 
an unstrained film (b) and when a strain gradient of  -1.53 $\times~10^7$~m$^{-1}$ (c) and 
-2.18$\times~10^7$~m$^{-1}$ (d) imposed in the zone delimited by a parallelepiped.
For all cases, the $3D$ map and its projection on the central $xy$ plane are shown. 
In (b) and (c) only one half of the supercell projection on that plane is plotted, for clarity, and 
the lines mark the position of the central vortices.
In (d), the line in the $3D$ map also indicates the positions of the central vortices,
and the projections on three planes are presented: 
(d,i) on the central $xy$ plane, where dashed lines indicate the positions
of two lateral planes, 
$L_x$ (d,ii) and $L_y$ (d,iii). 
Colors denote the $z$ components of the polarization.}
\label{fig:10c-grad}
\end{figure*}

Now we investigate how the PTO film responds to a compressive strain gradient
applied locally as its intensity increases. We are particularly interested in following the 
evolution of the domain configuration until eventually the polarization 
switches in the strain-gradient zone. As the first step, we have to choose where 
to apply the strain gradient. 
Based on the observed local strain distribution in the central 
region of each domain (Fig.~\ref{fig:latt_free}), we suppose that
a compressive strain gradient on the top surface of the film
would only accentuate the already existent inhomogeneous compression 
in the UP domain; on the contrary, it would alter the almost-constant 
$c$ distribution in the DOWN domain. Therefore, we apply to the latter 
a linear gradient on the top half of the film along the $z$ axis
as described in Sec. II and shown in Fig.~\ref{fig:10c-grad}(a) 
where the local $c$ profile is plotted before and after the application
for a particular $\eta_{zz}^{max}$ of -4.25~\%.
The same profile is established on a region of $4\times4$ unit cells
that extends from the top surface to the central plane~\cite{grad},
as schematically indicated in the figure by a parallelepiped. 
The strain gradient strength is increased progressively
by varying the compressive strain in the superficial cells 
$\eta_{zz}^{max}$ in intervals of -0.25~\%.
For each strain gradient value, we carry out an MD simulations until the system
rearranges to that strain condition.\\ \\
{\it Thinner films.} The results for the 10$c$ film are illustrated in 
Figs.~\ref{fig:10c-grad}(b)–~\ref{fig:10c-grad}(d) using three-dimensional (3D) maps of the local $P$ and 
projections of them on different planes. 
The unstrained film is shown in Fig.~\ref{fig:10c-grad}(b),
and how it changes when the imposed local strain gradient $g$ 
takes two particular values, 
namely, $g_1 =-1.53 \times 10^7$~m$^{-1}$ and $g_2 =-2.18 \times 10^7$~m$^{-1}$
in Figs.~\ref{fig:10c-grad}(c) and \ref{fig:10c-grad}(d), respectively.
Starting with the unstrained film,
which is repeated here so as to have a reference 
(same as the top panel of Fig.~\ref{fig:pol_free}), 
the 3D map displays a domain structure of 90$^\circ$ domain loops.
Its projection on the central $xy$ plane shows only two domains
with the $P$ downward and upward perpendicular to the paper plane 
separated by the central vortices, 
whose positions aligned along the $y$ axis are indicated by a line in the figure. 
The square in the DOWN domain identifies the region 
where the strain gradient will be applied. 
Thus, when the local strain gradient is imposed and as its strength $g$
increases, we find that the response of the film
presents three different behaviors. \\
(I) For $g < g_1$, no appreciable changes are observed 
and the local $P$ distribution remains similar to that of the unstrained film.\\
(II) For $g_1 \leq g < g_2$, when $g = g_1$, corresponding 
to an $\eta_{zz}^{max}$ of -3.00\%,
the whole domain structure is displaced, 
almost without changing either the size or the shape of the domains
as shown in the 3D map of Fig.~\ref{fig:10c-grad}(c).
The projection on the $xy$ plane clearly shows that
the central vortices have been shifted in $-x$ towards the gradient region 
and sit just before its border.
$P$ distributions similar to this one are observed
as the gradient is further increased.\\
(III) For $g = g_2$ or $\eta_{zz}^{max} =$ -4.25\%,
the dipoles rearrange and the local $P$s in the gradient region, 
initially downward, are switched 
[3D map of Fig.~\ref{fig:10c-grad}(d)]~\cite{inver}.
This region has merged with the original UP domain, 
extending its size and therefore changing its shape.
The new domain distribution can be appreciated more easily  
in the projection on the central $xy$ plane [Fig.~\ref{fig:10c-grad}(d,i)].
Also, we examine two lateral planes that cross the gradient region.
The projection on a $yz$ plane shown in Fig.~\ref{fig:10c-grad}(d,ii) 
(where all cells pointed downward before the gradient was applied)
reveals that the new domain pattern has developed a 
flux-closure arrangement around the central vortices analogous to 
that found in the unstrained film.
The projection on the $L_y$ plane [Fig.~\ref{fig:10c-grad}(d,iii)]
shows that most cells, including those in the gradient region, are poled upward.
The asymmetric distribution indicates that the 
switching process has started in the vortices 
that were brought close to the gradient region and now have disappeared, 
while the other vortices remain on the right side. \\
Note that due to the supercell in-plane periodicity,
it is equally likely that in (II) the vortices had shifted in $+x$ 
towards the gradient region and therefore the switching process 
would have started at the other side.
Further, we note that  the $c$ profile in the gradient region, whose inferior part 
[open circles in Fig.~\ref{fig:10c-grad}(a)] has been allowed to relax, 
has adopted a profile quite similar to the one observed in the UP domains
of the unstrained films, as was assumed initially.
Finally, we want to remark that the observed movements of domains 
in our free-standing films are allowed because of the absence of a substrate.
Its presence, however, can hamper the movement of domains,
as shown, for instance, in piezoelectric force microscopy images 
of Pb(Zr$_{0.2}$Ti$_{0.8}$)O$_3$ patterns in epitaxial thin films~\cite{naga02}. 

\begin{figure}
\begin{center}
\includegraphics[width=8.5cm,angle=0]{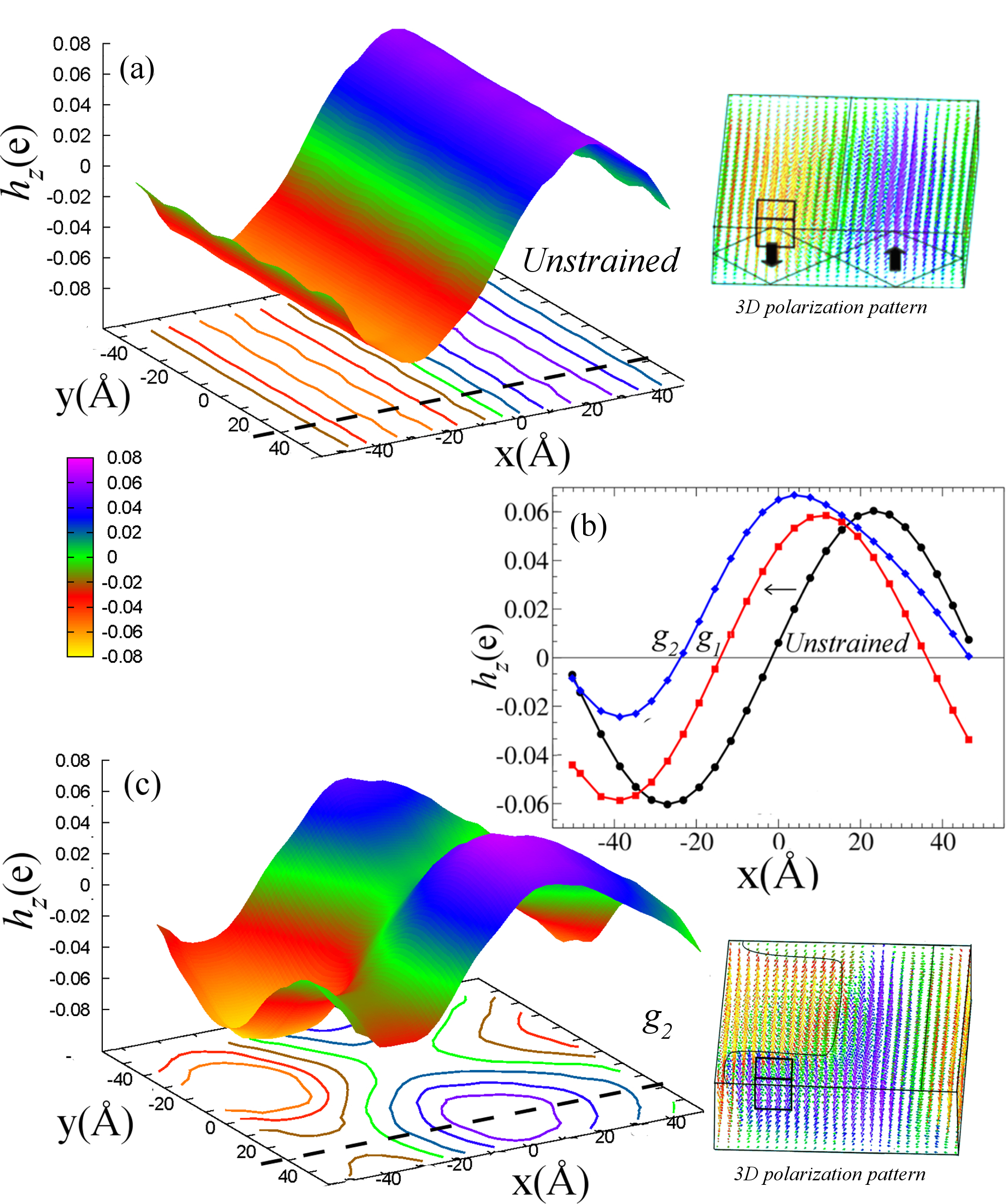}
\end{center}
\vspace{-15pt}
\caption{(Color online) Surface of the hypertoroidal moment $z$~component $h_z$
(in units of electronic charge $e$) 
as a function of the central $xy$ plane of the simulation cell,
for a 10$c$-thickness film 
(a) unstrained and (c) under a local strain gradient $g_2$.
For clarity, we include, to the right of each surface,
the corresponding $3D$ polarization pattern.
(b) Curves on a transveral $xz$ plane (indicated by dashed lines)
that crosses the strain gradient region.}
\label{fig:h_3D}
\end{figure}

{\it Hypertoroidal moment.}
In order to better describe the evolution of these complex dipolar configurations, 
we use the recently defined order parameter $\bf h$
named hypertoroidal moment~\cite{prosan07}, 
which provides a measure of subtle microscopic features. 
It involves the double cross product of the local dipoles $\bf p_{\it i}$ 
with the vectors ${\bf r}_{\it i}$ locating their positions,
\begin {equation}
{\bf h} = \frac{1}{4V} \sum_{i}{\bf r}_{\it i} \times ({\bf r}_{\it i} \times {\bf p}_{\it i_t})_{\it t}
\end {equation}
where $V$ denotes the supercell volume, 
the subindex $t$ labels the transversal component, and 
the summation runs over all unit cells within the supercell.
The transversal component can be approximated as 
${\bf p}_{\it i_t} \approx {\bf p}_{\it i}- \langle {\bf p} \rangle$,
$\langle {\bf p} \rangle$ being the average  of the 
individual dipoles over all the sites. This approximation is also valid for the 
expression in parenthesis.
We refer the reader to Ref.~\onlinecite{prosan07} for a detailed
discussion of this parameter. 
For 2D systems (i.e., systems periodic in two Cartesian directions 
and finite in the third one) such as the films considered here, 
once the simulation supercell is chosen,
the $\bf h$ value is independent of the origin used for $\bf r_{\it i}$.
However, the magnitude and even the sign of $\bf h$ do depend on the 
choice made for the supercell, leading to the multivaluedness of 
the $\bf h$ parameter.
We compute the hypertoroidal moment for different supercells
whose centers scan the complete central $xy$ plane of the original supercell.
Therefore we explore for the first time a surface of $\bf h$
and find that the changes in value and/or sign reflect 
modifications in the full $P$ distribution.

Although $P$ switching is a dynamical and complex process, 
we may gain some insight by examining the surface of only 
the $z$-component of the hypertoroidal moment $h_z$ 
(the other two Cartesian components are almost null) 
as the imposed strain gradient increases. 
Beginning with the unstrained film shown in Fig.~\ref{fig:h_3D}(a), 
we corroborate that the $h_z$ surface displays a sinusoidal dependence 
along the $x$ direction with the same period 
as the nanodomains and remains constant along the $y$ direction~\cite{prosan07}. 
Specifically, the maxima (minima) of $h_z$ extend along $y$, coinciding 
with the positions of the central cores of the UP (DOWN) domain, 
while $h_z$ vanishes when rotation around the perpendicular $y$ axis exists, 
certainly where the vortex centers sit. 
Then as the strain-gradient intensity increases and following the $P$ distributions, 
the $h_z$ response also presents three behaviors separated by 
the gradient values $g_1$ and $g_2$. \\
(I) For $g < g_1$, the sinusoidal shape remains almost unchanged. \\
(II) For $g_1 \leq g < g_2$, when $g = g_1$, in agreement with the $P$ distribution,
the $h_z$ surface shifts in $-x$ as shows by 
its projection on a transverse cut that crosses the central gradient region 
[see the central curve in Fig.~\ref{fig:h_3D}(b)].
$h_z$ behaves similarly as the gradient increases. \\
(III) For $g = g_2$, needed to fully invert the $P$ in the 
affected region, the $h_z$ surface shown in Fig.~\ref{fig:h_3D}(c) 
has changed dramatically, indicating that the 90$^\circ$-domain loop configuration 
has been modified. 
Two global extrema are observed, one maximum and one minimum, revealing that
the periodicity of the domain structure and thus the extended cores of the 
each domain have vanished.
From the $\bf h$ definition, we interpret that the $h_z$ minimum represents 
the part of the DOWN domain farthest away from the gradient region 
and therefore the less affected one, 
whereas the $h_z$ maximum is located in the region of the UP domain farthest away 
from the distribution of $-P_z$. 
Fig.~\ref{fig:h_3D}(b) shows that the $h_z$ profile in this particular direction
has slightly gained height in the positive branch, versus an important 
reduction in the negative one,
reflecting the size enhancement of the UP domain at expenses of the DOWN domain.\\
Summarizing, the evolution of the hypertoroidal moment as a function of the strain 
gradient strength provides information of the domain-structure evolution
localizing the domain cores and the vortices in a regular structure, 
as well as their redistribution when the $P$s in the strain region
are finally switched. 

\begin{figure}
\begin{center}
\includegraphics[width=8.cm,angle=0]{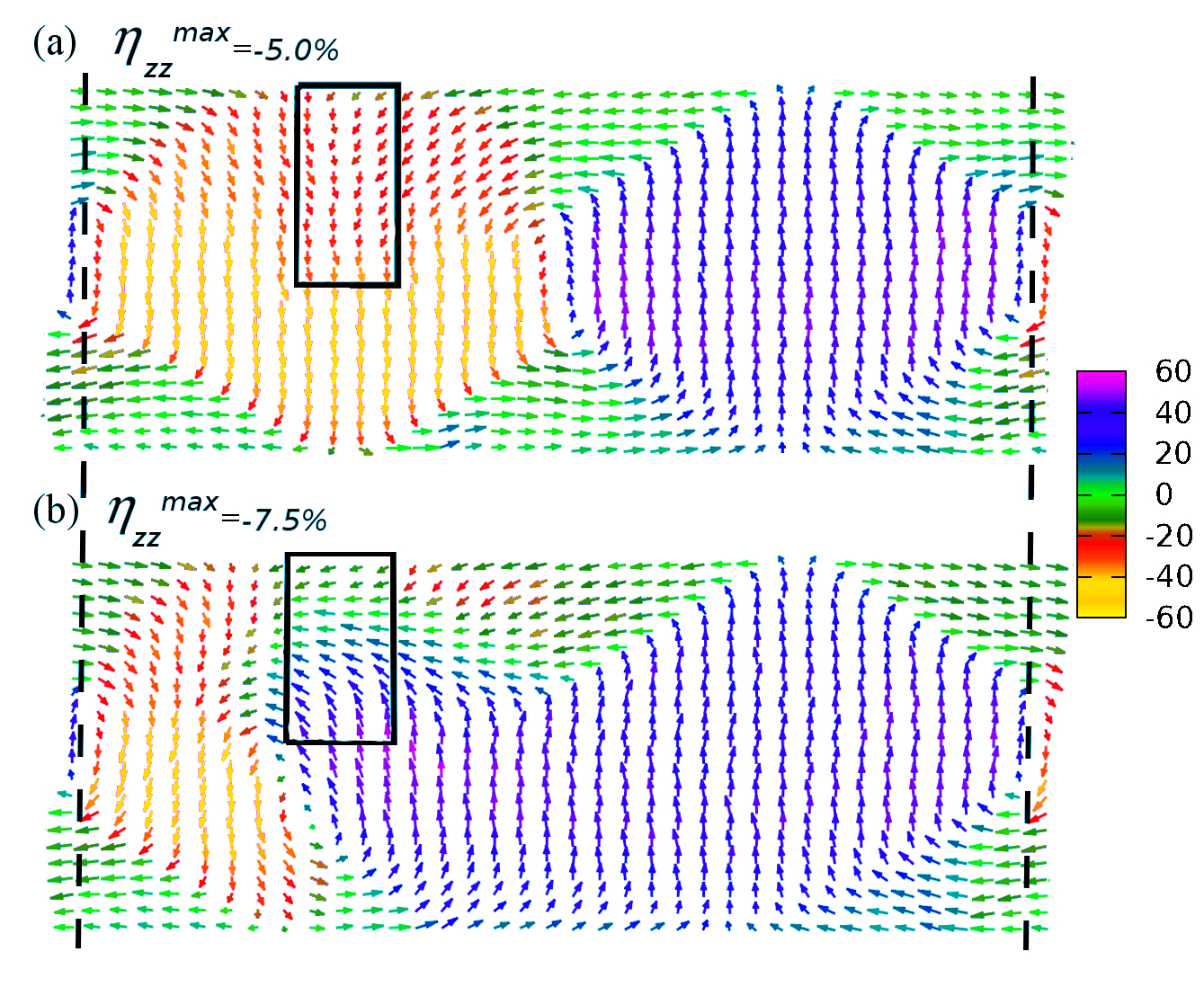}
\end{center}
\vspace{-15pt}
\caption{(Color online) 
Two-dimensional projection of the polarization patterns for an $xz$ plane that crosses the 
gradient region in a 23$c$-thickness film subject to 
a strain gradient of (a) $g_1^* =-1.01 \times 10^7$~m$^{-1}$ and (b) 
$g_2^* =-1.57 \times 10^7$~m$^{-1}$,
just before and after the switching, respectively.
The gradient region and the immobile 180$^\circ$ DW are indicated.}
\label{fig:grad_proy}
\end{figure}

{\it Thicker films.} 
Figure~\ref{fig:grad_proy} displays the $P$-map projection on a transverse cut 
crossing the gradient region of a $23c$-thickness film
for two strain-gradient values of $g_1^*= -1.01 \times 10^7$~m$^{-1}$ and $g_2^*= -1.57 \times 10^7$~m$^{-1}$,
corresponding to $\eta_{zz}^{max}$ of -5.00 and -7.50~\%, respectively.
As the gradient increases, starting at the surface 
the local $P$s near and inside the gradient region rotate progressively 
until pointing parallel to the surface, i.e., their $P_z$ components decrease gradually 
[as shown in Fig.~\ref{fig:grad_proy}(a) for $g_1^*$] until they fully vanish.
A lateral view in a $yz$ plane, not shown here, indicates that
the zone with $P$ parallel to the surface starts to grow
and expands towards the bottom of the film as the gradient increases,
then they are switched.
For the $g_2^*$ value [Fig.~\ref{fig:grad_proy}(b)] the affected region
is already poled upward and the asymmetric distribution shows
that one of the 180$^\circ$ DWs is still immobile.
Remarkably, as the gradient is increased further, 
the switched zone moves forward, generating vortices at the bottom
as it continues growing laterally.
Thus, unlike the DW movements observed in thinner films,
we find that the $180^o$ DWs never move forward to the gradient region.
This fact can be attributed to the relatively high energy barrier  
for the coherent motion of one entire $180^o$ DW 
compared to the $90^o$ DW case.\cite{mey02}

\vskip0.7cm

In summary, we investigate from an atomistic point of view flexoelectricity-induced
domain evolution. In free-standing PbTiO$_3$ ultrathin films 
with thickness-dependent domain patters, 
we study the ferroelectric switching induced by a strain gradient acting locally 
in the direction normal to the surface.
We first review the properties of strain-free films and 
confirm that the inhomogeneity of polarization, strain (strain gradient), and 
antiferrodistortive rotational angles appears not only among the surface cells 
but also in the neighborhood of the DWs. 
For the explored thicknesses (10 to 23 unit cells), we find flux-closure domain 
structures whose morphology is thickness dependent. 
A critical value of 20 unit cells is found: thinner films show structures 
with 90$^\circ$ domain loops, 
whereas thicker ones develop, in addition, $180^o$ DWs giving rise 
to structures of the Landau-Lifshitz type.  
When a compressive strain gradient normal to the surface acts locally at 
the top surface of the film, we find that it is able to 
reverse the normal polarization in DOWN domains but not in UP domains,
which is understood based on the knowledge of the local strain distribution 
before imposition of the gradient.
The evolution of the domain pattern as a function of the strain gradient strength 
depends, accordingly, on the film thickness.
In thinner films, first one of the center of the $90^o$ DW loops moves
towards the gradient region and then the affected region is gradually switched.
In thicker films, instead, no $180^o$ DW movement is observed
and the local polarizations in the gradient region 
are progressively reversed as the gradient strength is enhanced.
This work contributes from an atomistic point of view 
to the understanding of the mechanical flexoelectricity in tuning 
domains in ultrathin films.

\vskip0.7cm
The authors acknowledge support from the Argentinian Agency ANPCyT through Grants
PICT-PRH 99 and PICT-PRH 195.
G.D.B. thanks to CONICET-Argentina for the Ph.D. fellowship.


\begin{thebibliography}{999}

\bibitem{kog64}
S. M. Kogan, Sov. Phys. Solid State {\bf 5}, 2069 (1964).

\bibitem{zub13}
P. Zubko, G. Catalan, and A.K. Tagantsev, Annu. Rev. Mater. Res. {\bf 43}, 387 (2013).

\bibitem{yud13}
 P. V Yudin and a K. Tagantsev, Nanotechnology {\bf 24}, 432001 (2013).

\bibitem{lee12}
D. Lee, and T. W. Noh,
Phil. Trans. R. Soc. A {\bf 370}, 4944 (2012).

\bibitem{leePRL11}
D. Lee, A. Yoon, S. Y. Jang, J.-G. Yoon, J.-S. Chung, M. Kim, J. F. Scott, and T. W. Noh,
Phys. Rev. Lett. {\bf 107}, 057602 (2011).

\bibitem{gruv03}
 A. Gruverman, B.J. Rodriguez, a. I. Kingon, R.J. Nemanich, J.S. Cross, and M. Tsukada, 
Appl. Phys. Lett. {\bf 82}, 3071 (2003).

\bibitem{luSCI12}
H. Lu, C.-W. Bark, D. Esque de los Ojos, J. Alcala, C. B. Eom, G. Catalan, A. Gruverman,
Science {\bf 336}, 59 (2012).

\bibitem{fong04}
Fong, G.B. Stephenson, S.K. Streiﬀer, J. A. Eastman, O. Auciello, 
P. H. Fuoss, and C. Thompson, 
Science {\bf 304}, 1650 (2004).

\bibitem{mey02}
B. Meyer and D. Vanderbilt
Phys. Rev. B {\bf 65}, 104111 (2002)

\bibitem{bung05}
C. Bungaro and K. M. Rabe, Phys. Rev. B {\bf 71}, 035420 (2005).

\bibitem{sepli0506}
M. Sepliarsky, M. G. Stachiotti, and R. L. Migoni, 
Phys. Rev. B {\bf 72}, 014110 (2005),
Phys. Rev. Lett. {\bf 96}, 137603 (2006).

\bibitem{lai07}
B.-K. Lai, I. Ponomareva, I. Kornev, L. Bellaiche, and G. J. Salamo, 
Appl. Phys. Lett. {\bf 91}, 152909 (2007).

\bibitem{behe08}
R.K. Behera, B.B. Hinojosa, S.B. Sinnott, A. Asthagiri,
and S.R. Phillpot, J. Phys. Condens. Matter {\bf 20}, 395004 (2008).

\bibitem{shim08}
T. Shimada, K. Wakahara, Y. Umeno, and T. Kitamura,
J. Phys. Condens. Matter {\bf 20}, 325225 (2008).

\bibitem{shim10}
T. Shimada, S. Tomoda, and T. Kitamura,
Phys. Rev. B {\bf 81}, 144116 (2010).

\bibitem{qu13}
B. Yin and S. Qu, 
J. Appl. Phys. {\bf 114}, 063703 (2013).

\bibitem{ponoPRB12}
I. Ponomareva, A. K. Tagantsev, L. Bellaiche,
Phys. Rev. B {\bf 85}, 104101 (2012).

\bibitem{sepli04}
M. Sepliarsky, Z. Wu, A. Asthagiri, and R. E. Cohen,
Ferroelectrics {\bf 301}, 55 (2004).

\bibitem{dlpoly}
 W. Smith and T. R. Forester, computer code \textsc {DL$_-$POLY},
Daresbury and Rutherford Appleton Laboratory,
Daresbury, England.

\bibitem{wolf}
D. Wolf, P. Keblinski, S. R. Phillpot, and J. Eggebrecht, 
J. Chem. Phys. {\bf 110}, 8254 (1999).

\bibitem{poy99}
S. P\"{o}ykk\"{o} and D.J. Chadi, Appl. Phys. Lett. {\bf 75}, 2830 (1999); 
J. Phys. and Chem. of Solids {\bf 61}, 291 (2000).

\bibitem{belle}
G. Belletti and S. Tinte, unpublished (2014).

\bibitem{flex}
We have checked that the present atomistic model for PTO
is able to reproduce the longitudinal flexoelectric coefficient
in the range of $nC/m$.~\cite{belle}

\bibitem{DHV_11}
J. Hong and D. Vanderbilt, 
Phys. Rev. B {\bf 84}, 180101 (2011).

\bibitem{scott86} 
D. J. Srolovitz, and J. F. Scott, 
Phys. Rev. B {\bf 34}, 1815 (1986).

\bibitem{jia11} C.-L. Jia, K. W. Urban, M. Alexe, D. Hesse  and  I. Vrejoiu. 
Science  {\bf 331}, 1420 (2011).

\bibitem{wu06}
X. Wu and D. Vanderbilt, Phys. Rev. B {\bf 73}, 020103 (2006).

\bibitem{gur11}
M. Y. Gureev, A. K. Tagantsev, and N. Setter, Phys. Rev. B {\bf 83}, 184104 (2011).

\bibitem{junq12}
Pablo Aguado-Puente and Javier Junquera,
Phys. Rev. B {\bf 85}, 184105 (2012).

\bibitem{step01}
A. Munkholm, S.K. Streiffer, M.V. Ramana Murty, J.A. Eastman, C. Thompson, O. Auciello, L. Thompson, 
J.F. Moore, and G.B. Stephenson, 
Phys. Rev. Lett. {\bf 88}, 016101 (2001).

\bibitem{grad}
Two comments on the arbitrary size of the affected region: 
(i) that its deep depends on the film thickness indicates that,
for the same compressive strain on the superficial unit cell, 
the gradient is weaker in thicker films; (ii) 
that the lateral size is kept constant means that the 
affected region is farther away from the 180$^\circ$ DW
as the domain size increases together with the film thickness.

\bibitem{inver}
For a given value of strain gradient, the switching process ocurrs 
as a function of time during a molecular dynamics simulation,
noting that the switching kinetics is beyond the scope of this work.

\bibitem{naga02}
V. Nagarajan, A. Roytburd, A. Stanishevsky, S. Prasertchoung, T. Zhao, et al. 
Nat. Mater. {\bf 2}, 43–4 (2002).

\bibitem{prosan07} S. Prosandeev, L. Bellaiche, 
J. Mater. Sci. {\bf 44}, 5235 (2009).

\end{thebibliography}
\end{document}